\documentclass{article}
\usepackage{spconf,amsmath,graphicx}
\usepackage{algorithm}
\usepackage{algorithmic}
\usepackage{amsfonts}
\usepackage{url}
\usepackage{geometry}
\usepackage{graphicx}
\usepackage{float}
\usepackage{subfig}
\usepackage{multirow}
\usepackage{multicol}
\usepackage{mathtools}
\usepackage{booktabs}
\usepackage{subfig}
\usepackage{xcolor}
\usepackage{subfloat}
\title{DoRF: Doppler Radiance Fields for \\ Robust Human Activity Recognition Using Wi-Fi}
\name{Navid Hasanzadeh{$^{1}$} and Shahrokh Valaee$^{1}$, Fellow, IEEE}
\address{$^{1}$Department of Electrical \& Computer Engineering, University of Toronto, Toronto, ON, Canada\\{\textit{\small  navid.hasanzadeh@mail.utoronto.ca, valaee@ece.utoronto.ca.}}}
\begin{document}
%\ninept
%
\maketitle
\begin{abstract}
Wi-Fi Channel State Information (CSI) has gained increasing interest for remote sensing applications. Recent studies show that Doppler velocity projections extracted from CSI can enable human activity recognition (HAR) that is robust to environmental changes and generalizes to new users. However, despite these advances, generalizability still remains insufficient for practical deployment.
Inspired by neural radiance fields (NeRF), which learn a volumetric representation of a $3$D scene from $2$D images, this work proposes a novel approach to reconstruct an informative $3$D latent motion representation from one-dimensional Doppler velocity projections extracted from Wi-Fi CSI. The resulting latent representation is then used to construct a uniform Doppler radiance field (DoRF) of the motion, providing a comprehensive view of the performed activity and improving the robustness to environmental variability.
The results show that the proposed approach noticeably enhances the generalization accuracy of Wi-Fi-based HAR, highlighting the strong potential of DoRFs for practical sensing applications.

\end{abstract}
\section{Introduction}
\label{sec:intro}

Wi-Fi sensing has emerged as a practical approach for indoor human activity recognition and tracking, offering distinct advantages over traditional methods such as cameras and wearable devices~\cite{radwan2025tutorial}. Using existing Wi-Fi infrastructure eliminates the need for additional hardware, making deployment more accessible and cost-effective. Unlike wearable sensors, Wi-Fi sensing does not require individuals to carry or wear devices.
Additionally, unlike cameras, it does not record identifiable visual information, leading to improved privacy.

Wi-Fi-based HAR estimates human motion by analyzing how it alters the wireless channel, captured through Channel State Information (CSI), which reflects signal modifications due to reflections and scattering~\cite{salehinejad2022litehar,ding2024multiple}. Prior works have used either the magnitude or phase of CSI~\cite{yousefi2017survey, djogoHAR, jang2025study, zhang2024csi, peng2023rosefi}, often with machine learning. Magnitude-based methods work well in static settings but are sensitive to environmental changes, while phase-based methods~\cite{wu2022wifi, yin2022fewsense} suffer from issues like phase wrapping and hardware noise. Unfortunately, both approaches show limited accuracy and generalization, especially across users or environments~\cite{varga2024exposing}.

Alternative approaches extract Doppler velocity from Wi-Fi CSI, capturing frequency shifts caused by motion while suppressing static structures. Unlike most methods that estimate a single velocity by aggregating all multipath components~\cite{meneghello2022sharp, zhang2021widar3}, a recent work, MORIC~\cite{hasanzadeh2025moric}, extracts multiple Doppler projections, interpreted as views from virtual cameras distributed on a sphere. Each camera represents the net effect of multipaths modeled by a von Mises–Fisher distribution, enabling a more structured view of motion and bridging the gap between raw CSI and high-level activity features. This representation significantly improves generalization, especially for unseen users.
However, the random nature of environmental reflections causes the viewing angles of each WiFi access point (AP) to be limited and variable over time. 
These discrepancies hinder uniform coverage of the motion space and
make it challenging to synchronize observations across trials.
As a result, motion is observed from only a few scattered perspectives, similar to seeing a three-dimensional object through a few two-dimensional images.  When tested on new data representing unseen views, the model struggles to generalize due to the lack of a complete representation.

To overcome the challenges posed by the randomness and limited perspectives of Doppler projections, this paper introduces a method inspired by {\it neural radiance fields} (NeRF)~\cite{mildenhall2021nerf} to reconstruct a unified and informative three-dimensional latent representation of motion. By learning this representation from the one-dimensional Doppler velocity projections extracted from WiFi CSI, the method effectively aggregates partial and scattered views into a coherent global model of the activity. This latent representation is then used to generate a uniform {\it Doppler radiance field} (DoRF) that simulates observations from a complete set of virtual viewpoints, regardless of the actual AP configuration or environmental reflections. As a result, the approach mitigates the limitations of partial observations and enables classifiers to generalize more effectively to new and unseen data. The results show a noticeable improvement in Wi-Fi-based HAR generalization accuracy, confirming the effectiveness of the proposed approach.

\section{Delay-Doppler Decomposition of CSI}

% ---------- snippet starts ----------
\begin{figure*}[!t]
	\centering
	\subfloat{%
		\includegraphics[width=0.99\linewidth]{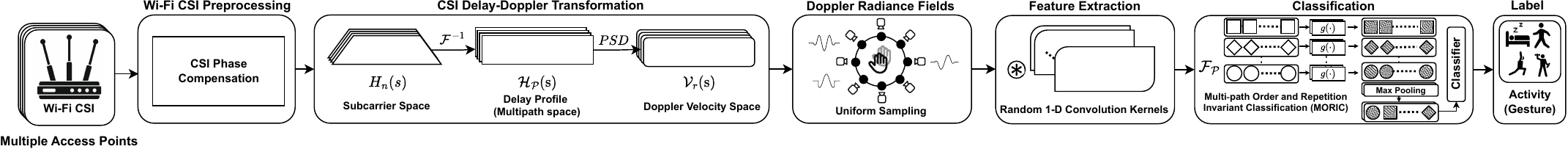}
	}
	\caption{The diagram illustrates the proposed Wi-Fi-based HAR framework built on MORIC. CSI signals are first converted into Doppler velocity projections that capture motion from multiple sparse spatial directions. These projections are then used to construct one or more Doppler radiance fields (DoRFs), which are uniformly sampled to generate multiview representations. A feature extraction module, followed by a classifier designed to be robust to multipath randomness, is used to recognize the performed activity.
}
\label{figure:method}
\end{figure*}
Consider a narrow-band Wi-Fi link with $N$ OFDM subcarriers of spacing $\Delta f$ and carrier frequency $f_c$. At time~$s$, the channel–state matrix at subcarrier~$n$ is modeled as  
\begin{align}
H_n(s) &= \sum_{l=0}^{L-1} \beta_l(s)\,
          e^{-j 2\pi f_n \,\tau_l(s)}, \label{eq:csi_f1}\\[4pt]
f_n     &= f_c - \Bigl(n-\tfrac{N}{2}\Bigr)\,\Delta f ,\label{eq:csi_f2}
\end{align}
where $H_n(s)$ is the complex CSI at subcarrier~$n$ and time~$s$; $\beta_l(s)$ and $\tau_l(s)$ are the complex gain and delay of the $l$-th multipath component; $f_n$ is the frequency of subcarrier~$n$; and $L$ is the number of multipath components.

An $N$-point inverse discrete Fourier transform (IDFT) across subcarriers yields the delay profile  
\begin{equation}
h(s,\tau)=\frac{1}{N}\sum_{n=0}^{N-1} H_n(s)\,
          e^{\,j\,2\pi n\,\Delta f\,\tau},\label{eq:idft}
\end{equation}
where $h(s,\tau)$ denotes the channel response at delay $\tau$ and time $s$.  
The IDFT effectively decomposes the frequency-domain CSI into a sum of delay-specific contributions,  
separating the effects of different multipath components based on their propagation delays.  
It aligns the phases of frequency components corresponding to the same excess delay, so that energy at  
\begin{equation}
\tau_i=\frac{i}{N\Delta f},\qquad i=0,1,\dots,N-1,\label{eq:taui}
\end{equation}
constructively adds, while energy elsewhere tends to cancel.  
Therefore, each resolvable bin $\tau_i$ can be interpreted as capturing the contribution of a distinct delay path in the multipath channel.

When the subject moves, the $i$\,th path component acquires an additional delay
\begin{equation}
%\Delta\tau_i=\frac{v(s)\!\cdot\!m_i}{c}\,t,
\Delta\tau_i=\frac{v(s)\cdot m_i}{c}\,t,
\label{eq:delta_tau}
\end{equation}
where $m_i\in\mathbb{R}^3$ is the mean arrival direction of the $i$-th multipath component with respect to the moving point,  
$v(s)$ is the spatial $3$D velocity vector during a short window centered at time $s$,  
$c$ is the speed of light, and $t$ is the elapsed motion interval.  
Hence, the term $h(s;\tau_i)$ gains a phase that varies in time with %$v(s)\!\cdot\!m_i$.
$v(s)\cdot m_i$.

\begin{figure}[!b]
	\centering
	\subfloat{%
		\includegraphics[width=0.83\linewidth]{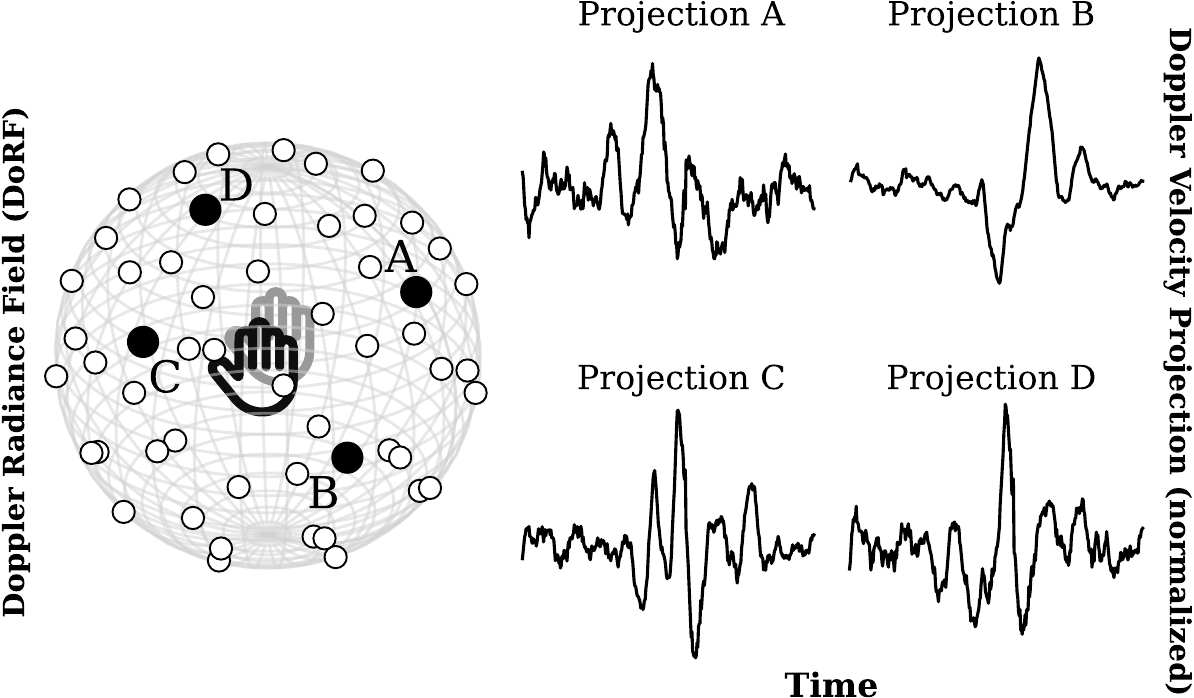}
	}
	\caption{Illustration of a DoRF constructed from Wi-Fi CSI. Sphere dots indicate observation directions, each yielding a one-dimensional Doppler projection. Four examples capture hand motion from different viewpoints. Uniform sampling ensures diverse coverage for robust motion representation.
}
\label{figure:Dorf}
\end{figure}

Intuitively, each direction $m_i$ can be interpreted as a viewpoint from which the motion is observed by the $i$-th multipath component. 
As shown in Fig.~\ref{figure:Dorf}, these directions act like virtual cameras surrounding the moving point, capturing the projection of the motion along different angles. 

Define the autocorrelation of the delay bin $\tau_i$ as
\begin{align}
R_h(\tau_i,t)
&=\mathbb{E}\!\bigl[h^{*}(s;\tau_i)\,h(s+t;\tau_i)\bigr],
\end{align}
and obtain the power spectral density by the Fourier transform
\begin{equation}
S(f;\tau_i)=\int_{-\infty}^{\infty} R_h(\tau_i,t)\,
e^{-j\,2\pi f t}\,dt.
\end{equation}
The spectrum attains its peak at the Doppler frequency  
$f^{\star}=v(s)\cdot m_i/\lambda$, with the wavelength $\lambda=c/f_c$.  
Converting this peak to a radial velocity yields
\begin{equation}
v_r\bigl(s;\tau_i\bigr)=\lambda\,f^{\star}=v(s)\!\cdot\!m_i.
\label{eq:vr}
\end{equation}

Collecting the projection from every resolvable delay bin forms the Doppler field
\[
\mathcal{V}_r(s)=
\bigl\{v_r(s;\tau_0),\,v_r(s;\tau_1),\dots,\,v_r(s;\tau_{N-1})\bigr\}.
\]
This set can be interpreted as simultaneous observations of the actual velocity from virtual cameras oriented in directions $\{m_i\}$ around the moving subject.
Although the exact directions $m_i$ are unknown and may repeat in arbitrary order, the ensemble $\mathcal{V}_r(s)$ still encodes the motion from diverse spatial viewpoints and provides a rich descriptor for subsequent activity classification.

\section{Method}
This section describes CSI phase pre-processing, the modeling framework for $3$D velocity estimation of the motion center of mass, the transformation into a uniform DoRF, and the activity classification method used in this work. Fig. \ref{figure:method} shows the different steps of the proposed method.

\subsection{CSI Phase Sanitization}

CSI phase measurements are distorted by hardware effects such as sampling frequency offset (SFO), symbol timing offset (STO), and random phase jumps, which can obscure motion-related variations. This work adopts the sanitization method of \cite{tadayon2019decimeter}, which unwraps the phase across subcarriers and fits a linear model to remove these dominant trends. The resulting phase-corrected CSI is then used for delay–Doppler processing.

\subsection{$3$D Velocity Estimation from Doppler Projections}

Let \( V_r \in \mathbb{R}^{T \times N} \) denote the matrix of Doppler velocity projections over \( T \) time steps and \( N \) paths. The entry at row \( s \) and column \( i \) is
\[
V_r(s,i) = v_r(s;\tau_i),
\]
where \( v_r(s;\tau_i) \) is the observed Doppler velocity for the delay bin \( \tau_i \).

The spatial $3$D velocity vector \( v(s) \in \mathbb{R}^3 \) is modeled as the integral of the actual local velocities in the environment over a short time window. This quantity approximates the instantaneous velocity of the aggregate center of mass of dynamic scatterers. Although it does not necessarily match the exact motion of a particular target point, such as the tip of a user's hand, it still serves as a robust descriptor of the overall action.

Each Doppler projection is expressed as
\[
v_r(s;\tau_i) = v(s)^\top r_i + n(s,i),
\]
where \( r_i \in \mathbb{S}^2 \) is an unknown unit direction and \( n(s,i) \) denotes measurement noise.  Stacking all observations over time and delay bins yields
\[
V_r = V R^\top + N,
\]
where \( V\in\mathbb{R}^{T\times3} \) contains the velocity vectors,  
\( R\in\mathbb{R}^{N\times3} \) the direction vectors,  
and \( N\in\mathbb{R}^{T\times N} \) the residual noise.

The estimation task is formulated as the following constrained optimization:
\begin{align}
\min_{V,R}\quad &
\frac{1}{2} \sum_{s=0}^{T-1} \sum_{i=0}^{N-1}
\left[ V_r(s,i) - v(s)^\top r_i \right]^2 \nonumber\\
&+ \lambda \sum_{s=0}^{T-1} \|v(s)\|^2
+ \gamma \sum_{i=0}^{N-1} \|r_i\|^2 \nonumber\\
\text{subject to} \quad &
\|r_i\| = 1, \quad i = 0,\dots,N{-}1.
\label{eq:objective}
\end{align}

Here, \( \lambda \) regularizes the magnitude of the velocity vectors to prevent overfitting, and \( \gamma \) improves stability during direction updates. Even though the constraint \( \|r_i\| = 1 \) is enforced by normalization, the \( \gamma \)-term further suppresses the effects of outliers and noise during least-squares updates. The full alternating optimization procedure for solving this non-convex problem is detailed in Algorithm~\ref{alg:alt-opt-3d-velocity}. In practice, dynamic time warping (DTW) is used to compare observed and predicted sequences over time, providing robustness to small misalignments. Additionally, a Procrustes alignment~\cite{ross2004procrustes} step is applied at each iteration to maintain consistency in the spatial orientation of the recovered motion basis.

\begin{algorithm}[!b]
\caption{Alternating Optimization for Estimating 3D Velocity of Motion Center of Mass}
\label{alg:alt-opt-3d-velocity}
\begin{algorithmic}[1]
\REQUIRE Doppler projection matrix \( V_r \in \mathbb{R}^{T \times N} \), tolerance threshold \( \epsilon \)
\STATE Initialize \( R = [r_0, \dots, r_{N-1}] \) with random unit vectors in \( \mathbb{R}^3 \)
\FOR{each iteration}
    \STATE \textbf{Velocity update:} for \( s = 0,\dots,T{-}1 \),
    \begin{equation}
    v(s) =
    \left( R^\top R + \lambda I_3 \right)^{-1}
    R^\top V_r(s,:)^\top
    \end{equation}
    
    \STATE \textbf{Direction update:} for \( i = 0,\dots,N{-}1 \),
    \begin{align}
    r_i &=
    \left( V^\top V + \gamma I_3 \right)^{-1}
    V^\top V_r(:,i) \\
    r_i &\leftarrow \frac{r_i}{\|r_i\|}
    \end{align}

    \STATE \textbf{Procrustes alignment:} apply rotation to \( V \) and \( R \) to stabilize orientation
    \STATE \textbf{Loss evaluation:} compute DTW-based loss between predicted and observed projections
    \IF{loss \( < \epsilon \)}
        \STATE \textbf{break}
    \ENDIF
\ENDFOR
\RETURN estimated velocity matrix \( V \), direction matrix \( R \)
\end{algorithmic}
\end{algorithm}

\subsection{Uniform Sampling of Doppler Radiance Fields}

Although the velocity sequence \( V \in \mathbb{R}^{T \times 3} \) provides a compact summary of motion, it lacks spatial diversity and is coordinate-dependent. To obtain a view-invariant representation, each velocity vector is projected onto a uniformly sampled grid of directions over the sphere.
Define the latitude and longitude angles for \( M \times 2M \) grid points:
\begin{align}
\theta_m &= \frac{(m+0.5)\pi}{M}, \quad m = 0, 1, \dots, M-1, \\
\phi_n &= \frac{(n+0.5)2\pi}{2M}, \quad n = 0, 1, \dots, 2M-1,
\end{align}
and construct the corresponding unit vectors:
\begin{equation}
d_{mn} =
\begin{bmatrix}
\sin\theta_m \cos\phi_n \\
\sin\theta_m \sin\phi_n \\
\cos\theta_m
\end{bmatrix}.
\end{equation}

For each time index \( s \), compute the radial projection:
\begin{equation}
P(s, m, n) = v(s)^\top d_{mn},
\end{equation}
which yields the DoRF
\begin{equation}
P \in \mathbb{R}^{T \times M \times 2M}.
\end{equation}

This radiance field captures instantaneous velocity components from all directions and provides a dense, structured representation of the performed activity. Unlike the raw Doppler projections, which are sparse and unordered, this view is complete and invariant to path geometry, making it highly suitable for downstream learning and classification.

\subsection{Activity Classification with MORIC}

The structured DoRF \( P \in \mathbb{R}^{T \times M \times 2M} \), obtained through uniform sampling over the sphere, is used as input to an activity classification model referred to as MORIC. This model is designed to handle sets of Doppler projections that may vary in number, density, and order depending on environmental conditions. Each Doppler projection is independently processed by a feature extractor which maps the projection into a high-dimensional feature space using random convolutional kernels and temporal pooling.

To ensure invariance to the ordering and repetition of projections, the MORIC classifier applies a max pooling operation across the projection axis. Let \( F_{s,i} \in \mathbb{R}^d \) denote the feature vector extracted from the \( i \)-th Doppler projection at time \( s \). Since the DoRF contains \( 2M^2 \) directional channels, \( i \in \{1, \dots, 2M^2\} \). The aggregated feature representation is computed as
\[
F_s = \max_{i=1,\dots,2M^2} F_{s,i},
\]
where the maximum is taken element-wise. This operation retains the most prominent features across the projections while eliminating redundant or noisy ones. Since the ordering and even the global rotation of Doppler directions on the sphere can vary across settings, max-pooling improves robustness to spatial variability.
The pooled feature \( F_s \) is then passed through a shallow classification network composed of fully connected layers and non-linear activations that outputs the class probabilities corresponding to the observed activity.

\section{Experiment}
\label{sec:pagestyle}
\subsection{Data}

\begin{figure}[!t]
	\centering
	\subfloat{%
		\includegraphics[width=0.9\linewidth]{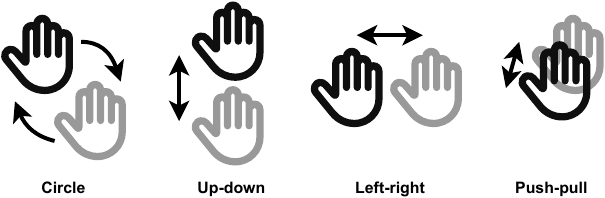}
	}
	\caption{Illustration of the four hand gestures performed by six participants in the UTHAMO dataset.
}
\label{figure:gestures}
\end{figure}

The UTHAMO dataset, introduced in \cite{uthamo}, is utilized to assess the performance of the proposed method. CSI data was collected from six participants performing four gestures—\textit{circle}, \textit{left–right}, \textit{up–down}, and \textit{push–pull}—using an ASUS RT-AC$86$U router equipped with three antennas in a static indoor office environment of size \(6\,\text{m} \times 5.6\,\text{m}\). The gestures are visually shown in Fig.~\ref{figure:gestures}. The gestures were performed between a Raspberry Pi that functions as a Wi-Fi transmitter at \(2.4\,\text{GHz}\) and the Wi-Fi router, placed directly in front of the user. CSI was acquired using the Nexmon toolkit~\cite{nexmon}, yielding \(52\) subcarriers per antenna. Each gesture was repeated for $20$ trials, each lasting five seconds, sampled at $100$~Hz.

\subsection{Training and Test Procedure}

The MORIC architecture was implemented in PyTorch using the default hyper-parameters and training setup described in \cite{hasanzadeh2025moric}. The configuration included $K=2$ attention heads, $D=1{,}000$ random kernels, a hidden layer size of $256$, and a reduced feature dimension of $D^{\prime}=128$. Training was performed using the AdamW optimizer~\cite{loshchilov2017decoupled} with a learning rate of $1\times10^{-4}$, a batch size of $64$, and cross-entropy loss with label smoothing ($\alpha=0.1$). A maximum of $2{,}500$ training epochs was used, with early stopping triggered after $200$ epochs without improvement in validation loss. A DoRF is individually constructed for each antenna of the AP and subsequently merged to form the input to MORIC. A uniform sampling grid of size $M=8$ is used and an $\varepsilon$ value of $0.01$ is used as the early stopping threshold in the alternating optimization algorithm.

To assess generalization performance, leave-one-subject-out (LOSO) cross-validation was used. In each iteration, data from one participant was reserved for testing, while the remaining data were used for training and validation. The model achieving the lowest validation loss was selected for final evaluation.

\subsection{Results}
\label{sec:typestyle}

Table~\ref{table:results} presents the generalization accuracies for four-class activity recognition on the UTHAMO dataset. Traditional CSI magnitude-based baselines, such as AMAP and CMAP, achieve performance close to the chance level, while more advanced Doppler and phase-based methods, including CapsHAR, CSI ratio, and APNSS$\,+\,$APSC, achieve modest improvements with accuracies ranging from $33\%$ to $39\%$.  
MORIC~\cite{hasanzadeh2025moric} significantly improves generalization performance, achieving a mean accuracy of $51.5\%$ using Doppler velocity projections. In comparison, the proposed method achieves the highest performance, with a mean accuracy of \textbf{54.8\%} and a reduced standard deviation of \textbf{7.2\%}.  
This improvement is attributed to the uniform sampling of DoRFs over the sphere, which provides comprehensive motion coverage and mitigates the effects of unknown viewpoints and multipath variability, thereby enhancing generalization to unseen users.

Conceptually, the proposed method shares similarities with unposed NeRFs~\cite{levy2023melon}, which reconstruct a $3$D scene by aggregating $2$D projections from multiple viewpoints with unknown positions. Analogously, this work builds a $3$D latent motion representation by aggregating one-dimensional Doppler velocity projections observed from multiple unknown directions. While NeRFs reconstruct spatial occupancy or color, DoRFs reconstruct directional motion velocities from sparse radial observations, yielding a motion field rich in temporal and spatial information. This representation not only improves classification, but also suggests potential for fine-grained motion reconstruction.

\begin{table}[!b]
\caption{Four-class hand motion generalization performance results averaged over six users.}
\small % reduce font size
\setlength{\tabcolsep}{5pt} % reduce horizontal padding
\begin{tabular*}{\linewidth}{@{\extracolsep{\fill}}lcc}
  \toprule
  \multirow{2}{*}{Method} &
    \multicolumn{2}{c}{Accuracy (\%)} \\
    & {Mean} & {Standard Deviation} \\
  \midrule
  AMAP \cite{salehinejad2023joint} & $27.4$ & $2.6$ \\
  CMAP \cite{salehinejad2023joint} & $27.5$ & $2.6$ \\
  CapsHAR \cite{djogoHAR} & $33.4$ & $4.2$ \\    
  CSI Ratio Model \cite{wu2022wifi} & $35.5$ & $5.6$ \\
  APNSS + APSC \cite{102859317} & $39.1$ & $6.0$ \\
  MORIC \cite{hasanzadeh2025moric} & $51.5$ & $8.7$ \\
  \hline
  \textbf{Proposed Method (DoRF)} & $\mathbf{54.8}$ & $\mathbf{7.2}$ \\
  \bottomrule
\end{tabular*}
\label{table:results}
\end{table}

\section{Conclusion}\label{section:conclusion}
This study introduces a novel Wi-Fi-based HAR method that constructs a unique motion representation by uniformly sampling the radiance fields extracted from CSI. Inspired by NeRF, it reconstructs a $3$D latent motion representation from one-dimensional Doppler velocity projections, capturing dynamic motion while suppressing static environmental effects. Projections are uniformly sampled over the sphere to enhance coverage and reduce sensitivity to viewpoint and multipath variability. The resulting projection fields, called DoRFs, provide a rich and directionally diverse representation that improves activity discrimination and robustness.
Experiments show that DoRF noticeably improves generalization to unseen users, outperforming prior methods in challenging cross-user settings, and demonstrating strong potential for practical Wi-Fi sensing applications.

\newpage

\bibliographystyle{IEEEbib}
\bibliography{refs}

\end{document}